\documentclass[
amsmath,
amssymb, 
twocolumn,
superscriptaddress,
a4paper,
11pt,
aps, 
prl]{revtex4-2}

\usepackage[T1]{fontenc}
\usepackage{newtxtext,newtxmath}

\usepackage{graphicx}
\usepackage{url}
\usepackage{hyperref}
\usepackage{lipsum}
\usepackage[margin=0.8in]{geometry}
 \hypersetup{
	colorlinks,
	linkcolor={blue!90!blue},
	citecolor={blue!10!blue},
	urlcolor={blue!80!blue}
}
\usepackage{xcolor}

\newcommand{\mstitle}{Length scale of cellular activity determines signatures of epithelial remodeling}

\newcommand{\sfI}{\textsf{I}}
\newcommand{\sfII}{\textsf{II}}
\newcommand{\sfIII}{\textsf{III}}
\newcommand{\sfIV}{\textsf{IV}}
\newcommand{\sfV}{\textsf{V}}

\begin{document}

	\title{\mstitle}
	
	% =========================================================
	% Authors and affiliations
	% =========================================================
	
	\author{Sahil~Islam}
	\email{ph22resch01009@iith.ac.in}
	\affiliation{
		Department of Physics, Indian Institute of Technology Hyderabad,
		Telangana, India
	}
	
	\author{Anupam~Gupta}
	\email{agupta@phy.iith.ac.in}
	\affiliation{
		Department of Physics, Indian Institute of Technology Hyderabad,
		Telangana, India
	}

	\author{Mohd.~Suhail~Rizvi}
	\email{suhailr@bme.iith.ac.in}
	\affiliation{
		Department of Biomedical Engineering,
		Indian Institute of Technology Hyderabad,
		Telangana, India
	}

	% =========================================================
	% Abstract
	% =========================================================
	
\begin{abstract}
	Cellular activity drives epithelial fluidization --- a widespread phenomenon observed during tissue development, remodeling, and repair both in vivo and in vitro.
	Yet the physical origins and spatial organization of active forces vary widely across biological systems and are often represented by a single generic mechanism in theoretical models.
	Here, using an active vertex model, we systematically compare four modes of epithelial activity spanning subcellular to tissue scales: apolar motility, polar motility, fluctuating contractility, and mechanochemical regulation.
	Although all four mechanisms drive the same global transition from a solid-like rectangular tissue to a fluid-like circular morphology, they reach this state through distinct pathways --- differing in the rates and topology of junctional rearrangements, cell elimination, and collective motion and leave distinguishable signatures in tissue architecture, cell dynamics, and mechanical relaxation.
	Among these observables, spatial velocity correlations directly capture the spatial organization of activity: their correlation length and functional form together resolve all four mechanisms.
	The robustness of these signatures across activity strengths suggests that spatial velocity correlations offer an experimentally accessible means of identifying the physical origin of epithelial activity from live-cell imaging alone.
\end{abstract}

\maketitle
	
	% \section*{Statement of significance}
	% Active forces are essential for epithelial tissues to change shape and reorganize, but the same tissue-scale behavior can emerge from activity generated at very different spatial scales.
	% Our results show that fluidization does not uniquely reflect the strength of cellular activity: the way forces are organized in space leaves distinct signatures in how cells rearrange, move, and dissipate mechanical stress.
	% In particular, spatial velocity correlations reveal these differences through characteristic correlation lengths and patterns of correlated and anti-correlated motion.
	% These findings show that collective cell motion can retain information about the spatial organization of active force generation, even when tissues exhibit similar large-scale behavior.
	
\section*{Introduction}

Epithelial tissues are among the most ubiquitous active materials in biology---forming the barriers, interfaces, and functional architectures of many organs \cite{guillot2013mechanics}.
Their ability to deform, flow, and remodel underlies diverse processes ranging from embryonic morphogenesis \cite{davidson2012epithelial} and wound healing \cite{pena2024cellular} to tissue homeostasis \cite{davidson2012epithelial,macara2014epithelial} and disease progression \cite{thiery2009epithelial,marchiando2010epithelial}.
These collective behaviors arise from the interplay between active and passive forces generated by individual cells through the actomyosin cytoskeleton and transmitted to neighbors through cell-cell adhesions \cite{guillot2013mechanics, vasquez2016force, kumar2025forces, heisenberg2013forces}, allowing tissues to actively reorganize while maintaining structural integrity.

A central manifestation of this activity is tissue fluidization \cite{guillot2013mechanics, kim2021embryonic, islam2025motility}.
Epithelia can transition from solid-like states that resist cell rearrangements to fluid-like states that readily reorganize, enabling large-scale shape changes and collective migration \cite{bi2015density, bi2016motility}.
Such transitions are essential during development, where tissues elongate, fold, and invaginate, and during wound repair, where coordinated cell rearrangements restore tissue integrity.
Similar mechanisms are implicated in pathological processes such as cancer invasion, where aberrant fluidization promotes tissue remodeling and dissemination \cite{marchiando2010epithelial, birchmeier1996epithelial}.

\begin{figure*}
	\centering
	\includegraphics[width=\linewidth]{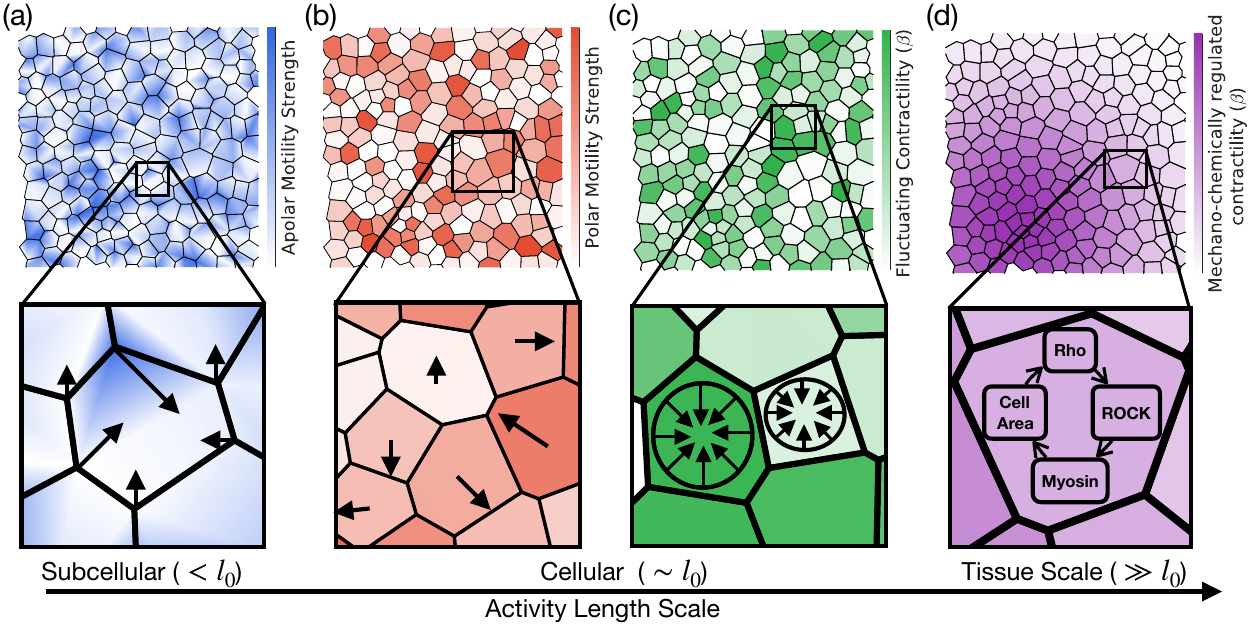}
	\caption{\textbf{Schematic representation of different classes of cellular activity.}\\(a) Apolar motility (AM): Activity originates from highly localized, stochastic forces (e.g., lamellipodia-like protrusions) acting at subcellular scales ($< l_0 (= \sqrt{A_0}) $ : cell size), resulting in temporally fluctuating, spatially uncorrelated forces at cell vertices with no directional bias.
		(b) Polar motility (PM): Each cell develops an internal polarity, generating coherent forces across its boundary along a preferred direction.
		The activity acts at the cell scale ($\sim l_0$).
		(c) Fluctuating contractility (FC): Activity arises from temporally varying, isotropic contractile stresses within each cell, reflecting stochastic actomyosin remodeling.
		The effective activity acts at the cellular scale but remains spatially uncorrelated across cells, with no directional bias ($\sim l_0$).
		(d) Mechanochemical regulation (MR): Contractility is governed by feedback between cell area and biochemical signaling pathways involving Rho–ROCK–myosin dynamics, leading to spatially and temporally correlated patterns of contractile activity. 
		This generates multi-cellular coordination ($\gg l_0$), often manifesting as propagating waves of contractility across the tissue.}
	\label{fig:schematic}
\end{figure*}

Although tissue fluidization is a common outcome, the underlying cellular mechanisms that can generate active forces are remarkably diverse.
Cells can actively deform their boundaries through stochastic protrusive fluctuations by cortical actin polymerization \cite{zimmermann2014formation, kumar2025forces}, migrate through persistent polarity-driven forces \cite{ridley2003cell, drubin1996origins}, remodel junctions via fluctuations in cortical contractility \cite{curran2017myosin}, or regulate force generation through mechanochemical feedback involving signaling pathways such as Rho--ROCK--myosin \cite{amano2010rho, koride2014mechanochemical}, or deploy a combination of these and similar other mechanisms.
These mechanisms differ fundamentally in the physical degrees of freedom they act upon and in the spatial and temporal scales over which activity is coordinated.
For instance, stochastic protrusive fluctuations originate from local rearrangements of the actin cortex at subcellular scales (Fig.~\ref{fig:schematic} (a)), whereas polarity-driven migration organizes forces across the scale of individual cells (Fig.~\ref{fig:schematic} (b)).
Cortical contractility fluctuations regulate cell shape and junctional tension through coordinated activity at cell-cell interfaces (Fig.~\ref{fig:schematic} (c)), while mechanochemical feedback can couple biochemical signaling and mechanical forces over multicellular length scales (Fig.~\ref{fig:schematic} (d)).
Together, these examples suggest that the spatial scale over which active forces are generated and coordinated may itself be a key determinant of how epithelial tissues remodel.

Despite this diversity in active force generation, most theoretical descriptions of active epithelia represent activity using a single generic mechanism, typically self-propelled motility \cite{bi2016motility, bi2015density} or an effective active stress \cite{kim2021embryonic}.
While such approaches successfully reproduce fluidization and collective motion, they implicitly assume that different sources of activity are mechanically equivalent.
Whether distinct force-generation mechanisms produce common tissue-scale behavior or instead leave unique structural, dynamical, and mechanical signatures remains largely unexplored.
Here, we seek to answer this question using four physically distinct classes of epithelial activity within a unified vertex-model framework that span different length scales of force generation: subcellular stochastic motility (activity length scale $<$ cell size, $l_0$) -- termed apolar motility (AM); cell-scale polarity-driven motion -- termed polar motility (PM) ($\sim l_0$); fluctuating cortical contractility (FC) ($\sim l_0$); and tissue-scale mechanochemical feedback, termed mechanochemical regulation (MR) ($\gg l_0$).

We observe that although all four classes drive tissue fluidization and large-scale remodeling, they do so through fundamentally different microscopic pathways.
We show that these differences are reflected in cell dynamics, tissue structure, and rheological responses.
Building on these observations, we identify experimentally accessible signatures that progressively distinguish the activity classes and demonstrate that spatial velocity correlations provide a unified fingerprint capable of resolving all four mechanisms.
Our results establish a physical framework for linking the scale and organization of active force generation to macroscopic tissue behavior, providing a practical route for inferring the dominant mode of cellular activity in epithelial tissues from live-imaging measurements.

\section*{Model}

We model a confluent epithelial tissue using a vertex-based framework \cite{honda2022vertex}, where each cell $c$ is represented as a polygon. 
The mechanical energy for the tissue is given by, 
\begin{equation}
	U = \sum_{c=1}^{N_c} \left[ \lambda (A_c - A_0)^2 + \beta P_c^2 + \gamma P_c \right],
	\label{eqn:energy}
\end{equation}
where $A_c$ and $P_c$ are the instantaneous area and perimeter of cell $c$, and $A_0$ is the preferred area.
The parameters $\lambda$, $\beta$, and $\gamma$ quantify cell elasticity, cortical contractility, and effective interfacial tension, respectively \cite{farhadifar2007influence}.
Vertex dynamics for a passive tissue follows the overdamped force balance equation,
\begin{equation}
	\eta \dot{\mathbf{r}}_{v} = -\nabla_v U,
	\label{eqn:force}
\end{equation}
where $\mathbf{r}_v$ is the vertex position and $\eta$ is an effective friction coefficient.
The gradient $\nabla_v$ is defined at the location of the vertex $v$.
Cells can undergo two types of topological rearrangements: Junctional rearrangements (T1 transition) and cell loss events (T2 transitions) (Fig.~\ref{fig:t1t2}).\\
The preferred cell area ($A_0$) sets a characteristic length $l_0 \sim \sqrt{A_0}$ and a single cell elastic relaxation timescale $\eta/(\lambda A_0)$.
Although we are not introducing new symbols, all mechanical parameters discussed further are non-dimensionalized using these scales.
For a typical epithelial cell, 
% standard length and timescales reads, 
$A_0 \sim 10^2$--$10^3\,\mu\mathrm{m}^2$, corresponds to cell sizes of $10$--$30\,\mu\mathrm{m}$ \cite{lin2018dynamic}, with area-relaxation timescales on the order of minutes ($\sim 100 ~\mathrm{s}$), set by the balance between cortical elasticity and dissipation \cite{lin2018dynamic}.

To investigate how the spatial scale and organization of active force generation influence epithelial remodeling, we extend the passive vertex model with the four activity classes introduced above.
These classes represent distinct physical mechanisms acting from subcellular to multicellular length scales while sharing the same underlying tissue mechanics. 
Their mathematical implementation and governing equations are described below.

\subsubsection*{Apolar Cell Motility (AM)}

Epithelial cells can generate active mechanical forces through stochastic protrusive fluctuations along their boundaries via the formation of filopodial extensions that rapidly extend and retract along the cell cortex \cite{zimmermann2013existence, zimmermann2014formation}.
These events produce highly localized, short-lived forces at discrete points along the cell perimeter, with no persistent direction \cite{ji2008fluctuations, zimmermann2014formation, machacek2006morphodynamic}.
Thus, their cumulative effect generates spatially uncorrelated and temporally fluctuating mechanical perturbations along the membrane (Fig.~\ref{fig:schematic}(a)).
We model this by augmenting the vertex dynamics with additive white noise felt at each vertex of a cell:
\begin{equation}
	\eta \dot{\mathbf{r}}_{v} = -\nabla_v U + \boldsymbol{\xi}_{v}(t),
	\label{eqn:force_apolar}
\end{equation}
with

\begin{equation}
	\resizebox{\linewidth}{!}{$
		\langle \xi^i_{v}(t) \rangle = 0, \qquad
		\langle \xi^i_{v}(t)\,\xi^j_{v'}(t') \rangle = 2\mathcal{M}_\mathrm{AM}\,\eta\,\delta_{vv'}\,\delta(t-t')\,\delta_{ij},
		$}
	\label{eqn:motility_AM}
\end{equation}
where $\mathcal{M}_{\mathrm{AM}}$ sets the motility amplitude, $i,j$ denote spatial components of $\boldsymbol{\xi}_{v}(t)$, and $v,v'$ denote vertices.

\subsubsection*{Polar Cell Motility (PM)}

Many epithelial cells establish an internal polarity axis through coordinated cytoskeletal organization: actin filament alignment, asymmetric myosin activity, and polarized trafficking of adhesion molecules \cite{ridley2003cell,petrie2009random, drubin1996origins}.
This bias stabilizes lamellipodia-like protrusions preferentially along one side of the cell, generating directed forces \cite{drubin1996origins}.
In the absence of strong external cues or long-range signaling, this polarity remains cell-intrinsic and spatially uncorrelated across the tissue (Fig.~\ref{fig:schematic}(b)).
We assign each cell $c$ an unit polarity vector $\hat{\mathbf{p}}_c (t) $ and apply a coherent active force to all its vertices:

\begin{equation}
	\eta \dot{\mathbf{r}}_{v} = -\nabla_v U + \,\hat{\mathbf{p}}_c (t),
	\label{eqn:force_polar}
\end{equation}
with polarity uncorrelated between cells,
\begin{equation}
	\left\langle p_c^i(t) \right\rangle = 0,
	\qquad
	\left\langle
	p_c^i(t)p_{c'}^j(t')
	\right\rangle
	=
	2\mathcal{M}_{\mathrm{PM}}\eta\,
	\delta_{cc'}\delta_{ij}\delta(t-t').
\end{equation}
with $i,j$ denoting spatial components of $\mathbf{p}_c$, and $c'$ denoting another cell.

\subsubsection*{Fluctuating Contractility (FC)}
Actomyosin contractility in epithelial cells is inherently dynamic, arising from stochastic myosin binding/unbinding and continuous actin cortex remodeling \cite{murrell2015forcing, curran2017myosin}. 
Consequently, junctional contractility exhibits both spatial heterogeneity between cell-cell interfaces and temporal fluctuations at individual junctions, even in the absence of a tissue-scale polarity \cite{curran2017myosin}.
These fluctuations generate isotropic, time-varying cortical tension rather than directed force generation (Fig.~\ref{fig:schematic}(c)). 
We model this behavior by assigning each cell $c$ a contractility $\beta_c(t)$ evolving as an Ornstein--Uhlenbeck process:
\begin{equation}
	\dot{\beta}_c = -\frac{\beta_c - \beta_0}{\tau_{\beta}} + \sigma_{\beta}\,\xi_c(t),
\end{equation}
where $\tau_\beta$ is the relaxation time, $\beta_0$ the reference contractility, and $\sigma_\beta$ the fluctuation amplitude.
The noise is Gaussian white noise:
\begin{equation}
	\langle \xi_c(t) \rangle = 0, \qquad
	\langle \xi_{c}(t)\,\xi_{c'}(t') \rangle = \delta_{{c}{c'}}\,\delta(t-t').
\end{equation}

\subsubsection*{Mechanochemical Regulation (MR)}

In many epithelia, contractility is not purely stochastic but is regulated by mechanosensitive signaling.
A central example is the Rho--ROCK pathway: mechanical deformation activates Rho GTPases, which drive downstream actomyosin contractility, thereby further deforming the cell \cite{koride2014mechanochemical, koride2018epithelial}.
This establishes a feedback loop coupling cell size to active stress generation.
We take the area deviation $\Delta A_c = A_c - A_0$ as the mechanical input driving sequential activation of Rho activity $\rho_c$, ROCK activity $R_c$, and myosin activity $M_c$ for a cell $c$ (Fig.~\ref{fig:schematic}(d)):
\begin{align}
	\dot{\rho}_c &= A_\rho\, h(\Delta A_c)\,\frac{(\Delta A_c)^n}{K^n + (\Delta A_c)^n}\,(1 - \rho_c) - D_A\,\rho_c, \\
	\dot{R}_c &= A_R\,\rho_c\,(1 - R_c) - D_R\,R_c, \\
	\dot{M}_c &= A_M\,R_c\,(1 - M_c) - D_M\,M_c,
\end{align}
where $h(\cdot)$ is the Heaviside function enforcing activation only under extension ($\Delta A_c \geq 0$), $K$ is the activation threshold, and $n$ controls cooperativity \cite{koride2014mechanochemical, koride2018epithelial}. 
The nonlinear activation terms ensure bounded, saturating dynamics.
The resulting contractility entering the mechanical energy is
\begin{equation}
	\beta_c(t) = \alpha\,M_c(t) + \sigma_\mathrm{MR}\,\xi_c(t),
\end{equation}
where $\alpha$ is the myosin--contractility coupling coefficient.
The noise term $\sigma_\mathrm{MR}\,\xi_c(t)$ captures fluctuations in this coupling, with $\xi_c$ Gaussian white noise satisfying
\begin{equation}
	\langle \xi_c(t) \rangle = 0, \qquad
	\langle \xi_{c}(t)\,\xi_{c'}(t') \rangle = \delta_{cc'}\,\delta(t - t'),
\end{equation}

Together, these four activity classes define a unified framework in which the passive mechanics of the tissue remain unchanged, while only the origin and spatial organization of active force generation are varied.
We next examine how these distinct mechanisms propagate across scales --- from topological rearrangements and cell dynamics to tissue mechanics and collective motion --- and identify experimentally accessible signatures of each activity class.

\begin{figure*}
	\centering
	\includegraphics[width=0.8\linewidth]{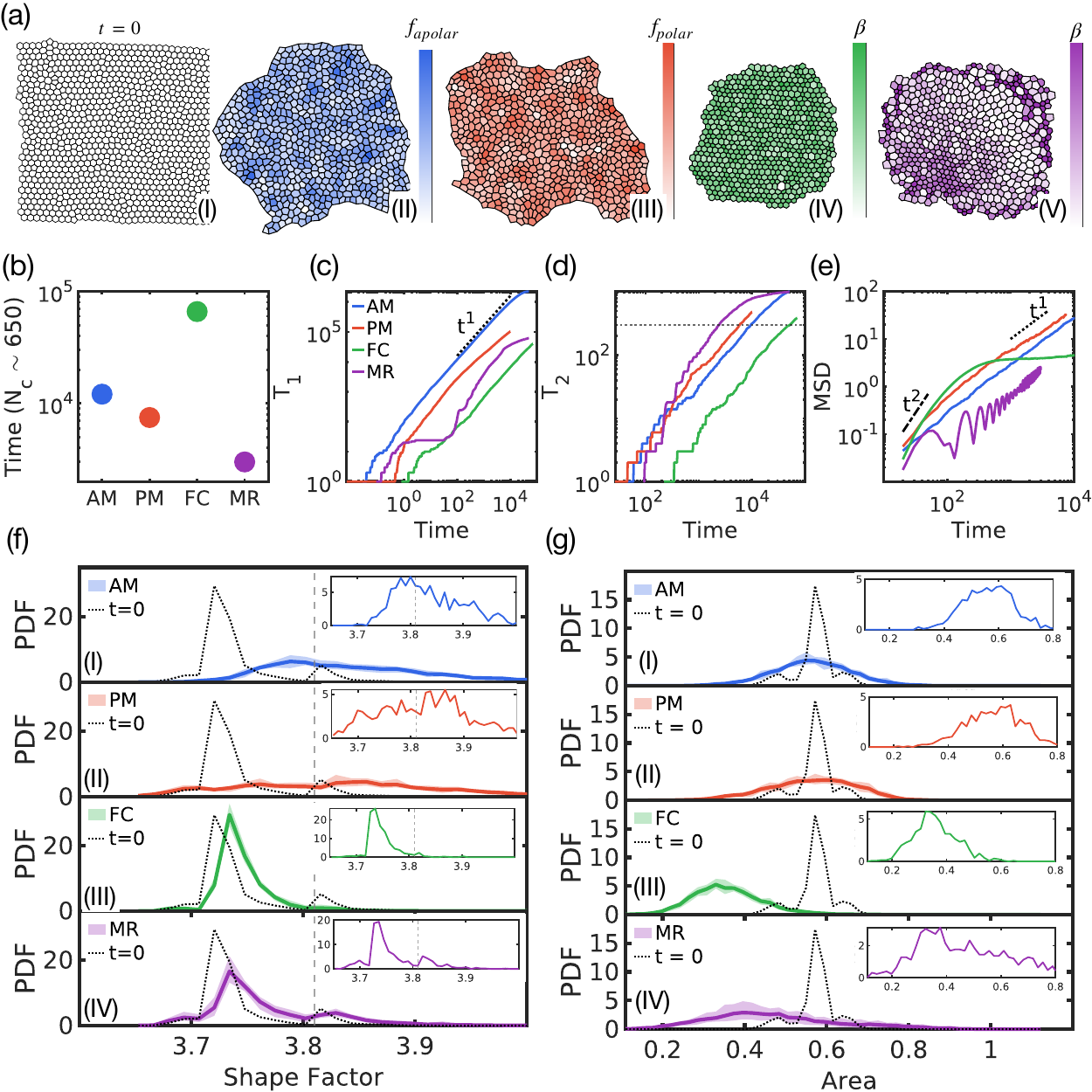}
	\caption{\textbf{Dynamical \& Structural analysis of activity-driven tissue fluidisation.}
		(a) (\sfI) Initial square epithelial tissue with free boundaries ($\sim10^3$ cells). 
		(\sfII)--(\sfV)). Final states for AM, PM, FC, and MR, respectively, all evolving toward a common circular reference state ($\sim650$ cells) for comparison. 
		Colorbars show the corresponding active field: stochastic forcing amplitude (AM, PM) or contractility (FC, MR).
		(b) Time to reach the reference state.
		(c) Cumulative T1 transitions.
		AM shows early and steady rearrangements due to direct vertex forcing; PM shows a delayed onset due to cell-scale coherent motion. 
		FC exhibits the slowest accumulation, reflecting weak edge-length fluctuations, whereas MR displays a two-stage response with an initial onset followed by rapid rearrangements driven by correlated contractility.
		(d) Cumulative T2 transitions.
		AM and PM show similar gradual accumulation; FC exhibits delayed and reduced cell loss. MR displays burst-like cell elimination followed by saturation, contributing to its rapid structural relaxation despite fewer T1 events.
		The dotted line shows $\sim 300$ T2 transitions. 
		(e) Mean square displacement (MSD).
		AM is diffusive; PM shows a weakly superdiffusive-to-diffusive crossover; FC exhibits an initial ballistic-like regime followed by caging; MR displays diffusion with damped oscillations from coordinated contractility dynamics.
		(f) Steady-state cell shape factor distributions.
		AM (\sfI) and PM (\sfII) show broadened distributions shifted toward anisotropic shapes, reflecting enhanced heterogeneity.
		FC (\sfIII) and MR (\sfIV) remain narrowly distributed below the rigidity threshold ($\sim3.81$), indicating fluidization without strong shape anisotropy.
		Plots in dotted lines show the distribution corresponding to the initial tissue state. 
		Vertical dashed line shows shape factor $=3.81$ --- the soft-to-hard transition value. 
		(g) Steady-state cell area distributions.
		AM (\sfI) and PM (\sfII) preserve the mean area but broaden the distribution through active deformation.
		FC (\sfIII) and MR (\sfIV) shift toward smaller areas due to contractility-driven shrinkage and T2-mediated cell loss, with MR showing additional broadening due to spatially correlated contractility.
		Insets in (f) and (g) show the corresponding distributions for the representative states in (a).
		Plots in dotted lines show the distribution corresponding to the initial tissue state.}
	\label{fig:circular}
\end{figure*}

\section*{Results}

\subsection*{Cellular activity drives tissue fluidization}

Cellular activity is known to promote large-scale shape remodeling and tissue fluidization in epithelia \cite{kim2021embryonic, islam2025motility}.
A natural manifestation of this is the spontaneous rounding of unconfined 
tissue into circular morphologies \textit{in vitro} \cite{michaut2025extracellular}.
We use this phenomenon as a common reference point to compare all four activity classes.
Starting from a rectangular tissue of $\sim 10^3$ cells under open boundary conditions, we investigate whether each class can drive this transition.

We find, under appropriate parameter regimes (see Methods for details), all four classes produce a circular morphology (Fig.~\ref{fig:circular}(a)) confirming fluidization.
To enable systematic comparison across classes, we define a common reference state: a circular tissue containing an equal number of cells.
%Simulation parameters are tuned so that each activity class reaches this reference state (see SI for details).
Despite converging to a similar final morphology, we observe that the time to reach the reference state differs across activity classes (Fig.~\ref{fig:circular}(b)).
AM and PM relax on comparable timescales, FC is substantially slower, whereas MR reaches the reference state most rapidly (Fig.~\ref{fig:circular}(b)).
This difference in rounding relaxation suggests that each activity class remodels the tissue through distinct mechanisms, naturally raising the question of how these mechanisms drive the topological rearrangements responsible for structural relaxation.

\subsubsection*{T1 transitions: onset and accumulation}

Large-scale epithelial remodeling is primarily mediated by junctional rearrangements (T1 transitions), driven by fluctuations in cell edge lengths \cite{farhadifar2007influence,kim2021embryonic, staple2010mechanics}.
Although all activity classes reach the same final state, their T1 dynamics differ substantially (Fig.~\ref{fig:circular}(c)).

AM exhibits the earliest T1 onset and a steady accumulation of rearrangements, as direct vertex forcing continuously generates edge-length fluctuations.
PM shows a delayed onset compared to AM because coherent cell-scale forces are initially insufficient to drive significant local edge asymmetry; rearrangements begin only after collective interactions generate sufficient imbalance.
Once initiated, PM accumulates T1 events at a rate comparable to AM, consistent with their similar relaxation timescales (Fig.~\ref{fig:circular}(c)).\\
In contractility-driven classes, edge-length fluctuations arise indirectly from spatial differences in cell tension.
FC exhibits the latest onset and the slowest accumulation, as spatiotemporally unstructured fluctuations in contractility generate only weak, transient differences in tension.
MR shows a distinct two-stage response: an initial increase in T1 activity is followed by a brief plateau until the Rho--ROCK--myosin feedback self-organizes into spatially coherent traveling waves (Fig.~\ref{fig:rho-rock-myosin}).
Once established, these waves periodically amplify tension differences between neighboring cells, producing coordinated edge-length fluctuations across the tissue.
Consequently, the cumulative T1 count increases with a much steeper slope than in other activity classes (Fig.~\ref{fig:circular}(c)), driving rapid junctional remodeling.
However, the total number of T1 transitions alone does not explain why MR exhibits the fastest structural relaxation, motivating an examination of additional topological processes, such as T2 transitions.

\subsubsection*{T2 transitions and the route to fluidization}

Under sustained contractility, cells can shrink below the stability threshold and be removed through T2 transitions, providing a distinct route toward structural remodeling (Fig.~\ref{fig:circular}(d)).
The contribution of this pathway depends strongly on the underlying mode of cellular activity.

AM and PM exhibit a gradual and nearly identical accumulation of T2 events, indicating that cell elimination accompanies remodeling at a steady rate, consistent with their similar relaxation dynamics.
FC produces very few T2 transitions, with delayed onset and slow accumulation.
Although contractility fluctuates markedly, its lack of spatial organization prevents coordinated, sustained cell shrinkage, thereby making elimination events rare.
MR displays a distinct response: after an initial accumulation comparable to AM and PM, the emergence of coherent Rho--ROCK--myosin waves drives a rapid burst of T2 transitions by collectively reducing cell areas below the critical threshold on timescales shorter than area relaxation.

These distinctions reveal that fluidization is governed not only by the frequency of junctional rearrangements but also by the efficiency with which cells are removed from the network.
Although MR does not produce the largest total number of T1 transitions at long times, its transient burst of T2 events provides an additional, irreversible mechanism of topological reorganization that rapidly accelerates tissue remodeling.
Conversely, FC remains the slowest-remodeling state because it is ineffective in both pathways.
The origin of cellular activity, therefore, determines not only the rate of fluidization but also the topological pathway through which it proceeds.

\subsection*{Dynamical signatures: Mean squared displacement}

The distinct T1 and T2 dynamics described above explain why the four activity classes fluidize on different timescales.
We next ask whether these mechanistic differences produce experimentally measurable dynamical signatures.
As a first dynamical measure, we examine the mean-square displacement (MSD), which quantifies cell transport during fluidization (Fig.~\ref{fig:circular}(e)).

AM exhibits diffusive scaling (MSD $\propto t^1$) across all timescales, as spatially uncorrelated vertex-level forcing continuously perturbs cell positions without directional persistence.
PM similarly exhibits diffusive behavior, but with a higher MSD at short and intermediate times.
The coherent cell-scale forcing generated by polar activity collectively shifts the vertices of each cell, producing larger initial cell displacements than the uncorrelated vertex-level forcing in AM.
However, because polarity directions are uncorrelated between neighboring cells, this cell-scale coherence does not produce persistent tissue-scale directional motion, resulting in diffusive transport at longer timescales (Fig.~\ref{fig:circular}(e)).\\
FC displays a distinct initial ballistic-like regime followed by a pronounced plateau.
The initial increase arises from the transient mechanical readjustment of the equilibrated tissue to the newly introduced contractility fluctuations, as cells move away from the initial configuration toward configurations favored by the instantaneous contractility.
As the fluctuating contractility continues to drive local displacement, these displacements remain spatially confined, leading to caging at $\sim2$ cell lengths rather than sustained cell transport (Fig.~\ref{fig:circular}(e)).\\
MR exhibits diffusive motion with superimposed damped oscillations arising from the traveling Rho--ROCK--myosin waves generated by the mechanochemical feedback loop (Fig.~\ref{fig:rho-rock-myosin}).
The emergence of spatially coordinated contractility periodically drives collective contraction and expansion of neighboring cell groups, producing oscillatory signatures in the MSD.
The damping of these oscillations arises from both irreversible tissue remodeling through T1 and T2 transitions and the loss of phase coherence between locally generated contractile waves due to stochastic fluctuations in the mechanochemical feedback.

The MSD therefore provides a partial dynamical fingerprint of activity type.
It distinguishes the localized, caged dynamics of FC from the transport-dominated AM, PM, and MR classes.
However, AM, PM, and MR all exhibit long-time diffusive behavior despite their fundamentally different mechanisms, highlighting that MSD alone cannot resolve all activity classes and motivating complementary observables based on the spatial organization of cellular motion.

\subsection*{Structural heterogeneity}

We next investigate whether tissue architecture alone can distinguish between different modes of cellular activity. 
To this end, we analyze the distributions of the cell shape factor ($\mathrm{Perimeter}/\sqrt{\mathrm{Area}}$) and cell area (Fig.~\ref{fig:circular}(f,g)), two geometric descriptors that are readily measurable from experimental images.

\subsubsection*{Cell shape distributions}

We begin from a nearly homogeneous tissue (Fig.~\ref{fig:circular}(a)) with an isotropic cell population, characterized by a mean shape factor below $\sim3.81$. 
We then let the tissue evolve until it reaches a steady fluidized state, and we compare the resulting shape factor distributions (Fig.~\ref{fig:circular}(f), insets).

Both motility-driven activity classes (AM and PM) shift the distribution toward larger shape factors while substantially broadening it, yielding mean values exceeding $\sim3.81$.
In AM, spatially uncorrelated vertex forces continuously deform individual cells, generating large shape fluctuations (Fig.~\ref{fig:circular}(f, \sfI).
In PM, coherent forces within each cell promote transient cell elongation, while the continual spatiotemporal reorientation of polarity broadens the distribution. 
The representative tissue snapshots (insets) clearly reflect these elongated, heterogeneous cell shapes (Fig.~\ref{fig:circular}(f, \sfII).

In contrast, the contractility-driven classes (FC and MR) retain mean shape factors below $\sim 3.81$, despite undergoing complete tissue fluidization (Fig.~\ref{fig:circular}(f, \sfIII, \sfIV).
Rather than producing persistent cell elongation, active tension fluctuations remodel the tissue primarily through junctional rearrangements.
MR exhibits a modestly broader distribution than FC because the propagating Rho--ROCK--myosin waves generate transient, spatially localized shape deformations as they sweep across the tissue, consistent with the representative snapshots shown in the insets (Fig.~\ref{fig:circular}(f, \sfIV).

These observations establish that pronounced cell shape anisotropy is not a prerequisite for epithelial fluidization.
While motility-driven activity fluidizes the tissue through sustained cell deformation, contractility-driven activity achieves the same through dynamic regulation of junctional tensions and topological remodeling.

\subsubsection*{Cell area distributions}

Cell area distributions provide a complementary structural fingerprint of the four activity classes (Fig.~\ref{fig:circular}(g), insets). 

The motility-driven classes (AM and PM) preserve the mean cell area while substantially broadening the distribution.
In AM, uncorrelated vertex forces continuously drive local expansion and compression, whereas in PM, polarity-driven directed forces generate comparable area heterogeneity (Fig.~\ref{fig:circular}(g, \sfI, \sfII).
The representative snapshots (insets) reflect these pronounced cell-to-cell variations in area.

The contractility-driven classes exhibit a qualitatively different signature.
Both FC and MR shift the distribution toward smaller areas, consistent with sustained contractility and the accompanying T2-mediated elimination of highly compressed cells.
FC retains a relatively narrow distribution, indicating that spatially uncorrelated contractility shrinks cells fairly uniformly across the tissue with open boundaries
(Fig.~\ref{fig:circular}(g, \sfIII).
In contrast, MR displays a broader distribution, as propagating Rho--ROCK--myosin waves create transient regions of enhanced contraction interspersed with less-contractile regions, thereby increasing spatial heterogeneity in cell area (Fig.~\ref{fig:circular}(g, \sfIV).

Taken together, the shape factor and area distributions reveal that motility-driven activity is characterized by enhanced cell-shape anisotropy and broad cell-area fluctuations with little change in mean area, whereas contractility-driven activity preserves nearly isotropic cell shapes while shifting the tissue toward smaller cell areas.
While these structural descriptors clearly distinguish the two broad classes of activity, further investigation is needed to distinguish between the individual activity modes within each class.

\begin{figure*}
	\centering
	\includegraphics[width=0.75\linewidth]{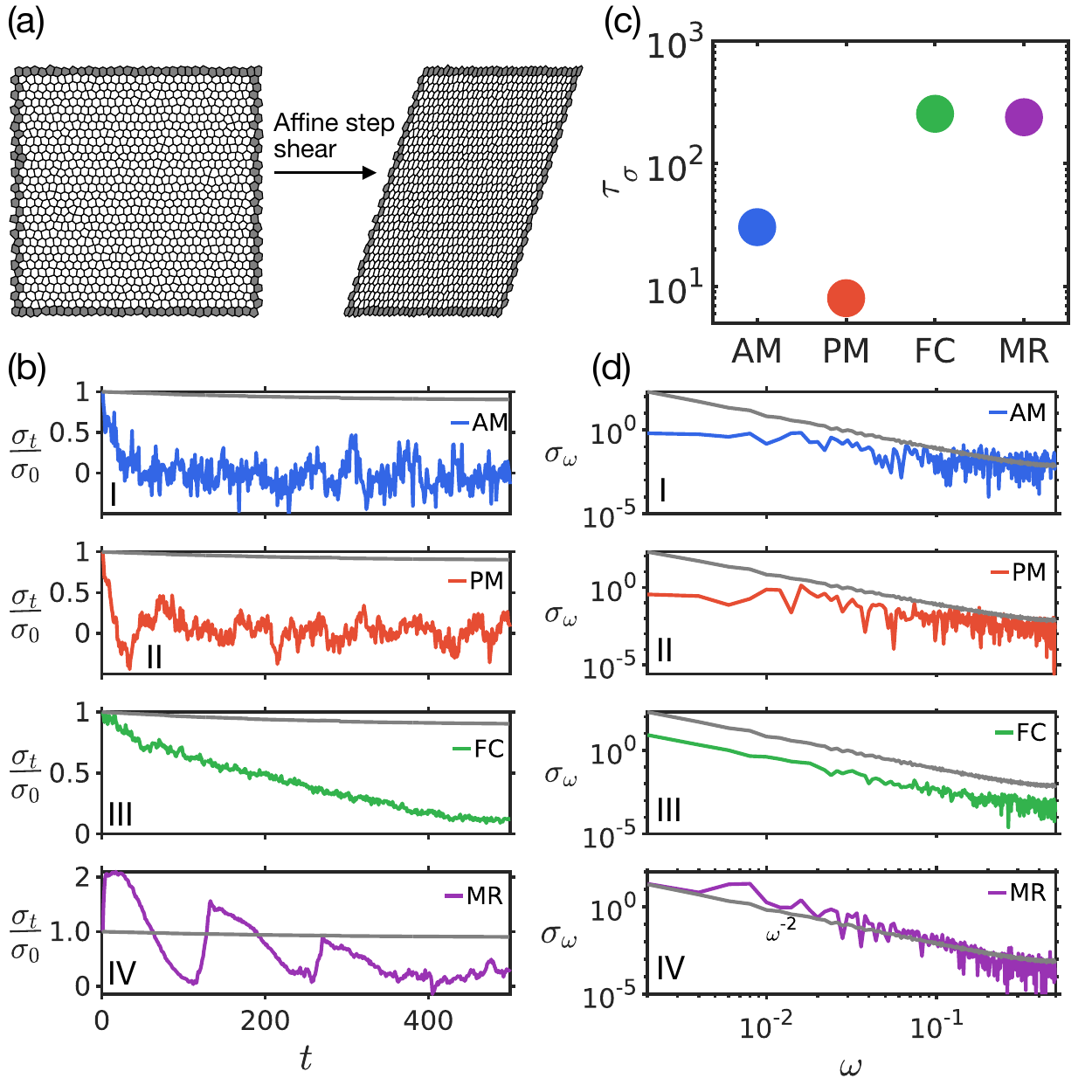}
	\caption{\textbf{Temporal and spectral signatures of stress relaxation across activity classes.}
		(a) Schematic of affine simple shear deformation.
		A $10\%$ strain is applied and held fixed, while bulk shear stress is measured away from the boundary.
		(b) Stress relaxation dynamics for different activity classes.
		AM (\sfI) and PM (\sfII) exhibit rapid relaxation followed by fluctuations around zero, indicating efficient dissipation of stress through force-driven rearrangements.
		FC (\sfIII) relaxes more slowly, as spatially unstructured contractility fluctuations generate local stresses that dissipate only through slow stochastic rearrangements.
		MR (\sfIV) exhibits non-monotonic, oscillatory relaxation driven by temporally correlated Rho--ROCK--myosin feedback, which periodically builds and releases tissue stress.
		Grey curves denote the corresponding relaxation of non-active tissue.
		(c) Characteristic relaxation times extracted from fits to the stress decay (see S.I. for details).
		AM and PM exhibit shorter timescales, whereas FC and MR display slower relaxation due to the indirect and temporally structured nature of contractility-driven activity. 
		(d) Fourier spectra of stress relaxation.
		Non-active tissue exhibits $\omega^{-2}$ scaling, consistent with exponential relaxation (amplitude rescaled for comparison).
		All active cases retain $\omega^{-2}$ behavior at intermediate-to-high frequencies, with distinct low-frequency signatures.
		AM (\sfI) and PM (\sfII) show a low-frequency plateau associated with loss of long-time stress correlations through sustained rearrangements.
		FC (\sfIII) preserves $\omega^{-2}$ scaling across frequencies, reflecting consistent exponential decay.
		MR (\sfIV) exhibits a pronounced spectral peak at $\omega\sim10^{-2}$, directly capturing the characteristic frequency of mechanochemical oscillations.}
	\label{fig:stress}
\end{figure*}

\subsection*{Rheological characteristics}

The structural differences identified above naturally raise the question of whether they are accompanied by distinct mechanical responses.
A widely used experimental approach to probe tissue mechanics is the stress relaxation test \cite{kim2021embryonic,papafilippou2025interplay}, which quantifies how rapidly an epithelial layer dissipates an externally imposed deformation.
We therefore subject the tissue to an affine simple shear deformation of $\sim 10\%$, hold the boundary fixed to keep the strain, and monitor the subsequent relaxation of the bulk shear stress (Fig.~\ref{fig:stress}(a)).
We characterize the response through the time evolution of the global shear stress (Fig.~\ref{fig:stress}(b)) (see Methods for definition), its characteristic relaxation timescale (Fig.~\ref{fig:stress}(c)), and the Fourier spectrum of stress fluctuations during relaxation (Fig.~\ref{fig:stress}(d)).

\subsubsection*{Stress relaxation: time evolution}

AM and PM both exhibit rapid, monotonic decay of shear stress, followed by low-amplitude fluctuations around zero at long times (Fig.~\ref{fig:stress}(b, \sfI \& \sfII)). 
This is consistent with the efficient, force-driven rearrangements established above: continuous vertex-level forcing in AM and coherent cell-scale polar forcing in PM generate sustained junctional rearrangements that rapidly dissipate built-up stress. 
In contrast, the passive tissue relaxes much more slowly, as stress release relies solely on mechanically driven rearrangements in the absence of active remodeling
(Fig.~\ref{fig:stress}(b, grey curves).
The residual fluctuations in AM and PM reflect the ongoing but uncoordinated nature of active forcing once the tissue has fluidized.\\
For FC, stress is released primarily through intermittent T1 and T2 events driven by gradual contractility fluctuations. 
This indirect coupling between activity and rearrangements suppresses rapid stress release, resulting in relaxation timescales longer than AM and PM but still shorter than the passive case (Fig.~\ref{fig:stress}(c, \sfIII)).\\
MR exhibits a damped oscillatory relaxation (Fig.~\ref{fig:stress}(b, \sfIV)). 
The area--Rho--ROCK--myosin mechanochemical feedback periodically amplifies and relaxes contractility in a spatially coordinated manner, producing repeated cycles of stress buildup and release superimposed on the overall relaxation. 
The oscillation amplitude progressively decreases because irreversible T1 and T2 rearrangements continuously dissipate the elastic stress imposed by the initial shear, leaving less stored stress for subsequent feedback-driven modulation.
Consequently, the relaxation envelope decays more slowly than in AM and PM (Fig.~\ref{fig:stress}(b)), yielding a larger characteristic relaxation time (Fig.~\ref{fig:stress}(c)).

\subsubsection*{Stress relaxation: Fourier spectra}

The Fourier spectra of stress relaxation provide an additional fingerprint of the activity classes (Fig.~\ref{fig:stress}(d)).
The passive tissue exhibits $\omega^{-2}$ scaling over the measured range, consistent with exponential stress relaxation (Fig.~\ref{fig:stress}(d), grey curves).
All active cases retain this scaling at high frequencies, reflecting the generic finite-timescale relaxation of an overdamped system.

The low-frequency behavior distinguishes the activity classes.
AM and PM develop a low-frequency plateau, indicating loss of long-time stress memory due to sustained rearrangements.
FC maintains the $\omega^{-2}$ scaling across the full frequency range, consistent with slow, rearrangement-limited relaxation.
MR exhibits a distinct spectral peak at $\omega\sim10^{-2}$ superimposed on the $\omega^{-2}$ background, directly capturing the characteristic frequency of Rho--ROCK--myosin oscillations and the associated cycles of stress buildup and release.
Importantly, the spectral representation reveals the underlying relaxation dynamics even when the stress decay is weak within the observation window, as demonstrated in the passive case.

Together, stress relaxation and its spectral signatures distinguish the contractility-driven classes, with the MR peak providing a distinctive signature of mechanochemical feedback.
Such signatures can be accessed experimentally through stress-relaxation rheology \cite{safa2024active,papafilippou2026emergent}, microrheology \cite{janshoff2021viscoelastic}, or traction force microscopy \cite{style2014traction} combined with temporal spectral analysis.

However, AM and PM remain difficult to distinguish using the structural and rheological observables considered thus far.
Because their primary difference lies in the spatial organization of active forces rather than in their bulk structural or mechanical response, we hypothesize that spatial velocity correlations among cells could encode the missing information needed to distinguish all four activity classes.

\subsection*{Cellular velocity correlation resolves all activity classes}

To test this hypothesis, we compute the spatial velocity correlation function, which quantifies the spatial coordination of cellular motion (see Methods section for mathematical details).
The resulting correlation profiles reveal how cell movements are organized across different length scales and provide an experimentally accessible measure of the underlying mechanism of activity.

\begin{figure}
	\centering
	\includegraphics[width=\linewidth]{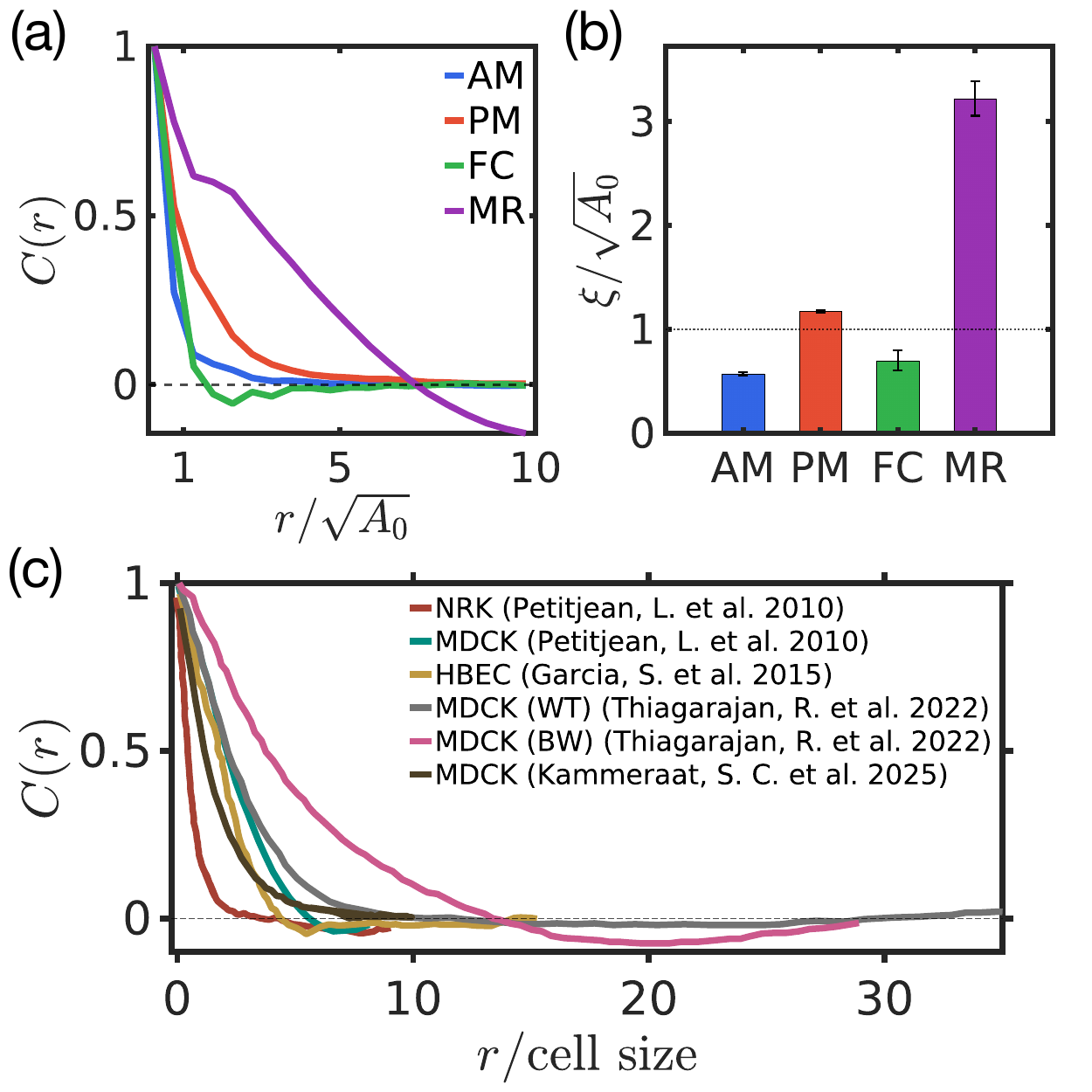}
	\caption{\textbf{Spatial velocity correlation across activity classes.}
		(a) Spatial velocity correlation $C(r)$ as a function of normalized inter-cellular distance ($r/\sqrt{A_0}$). 
		AM and PM exhibit exponential decay, with PM showing an extended correlation length due to polarity-induced coherent motion. 
		FC displays rapid decorrelation followed by a shallow negative lobe, reflecting local anti-alignment generated by neighboring contractile and extensile fluctuations. 
		MR shows the slowest decay and a pronounced negative lobe, arising from mechanochemical waves in which cells within a wavefront share similar contractile states, whereas cells separated by approximately one wavelength undergo opposite motions.
		(b) Correlation lengths $\xi/\sqrt{A_0}$ extracted from fits to the curves in (a). 
		AM and FC remain correlated over sub-cellular scales, PM shows cell-scale coherence, and MR exhibits the largest correlation length ($\sim3\sqrt{A_0}$).
		(c) Experimental measurements of spatial velocity correlations in different epithelial tissues (MDCK, NRK, and HBEC) from different studies \cite{petitjean2010velocity,garcia2015physics,thiagarajan2022pulsations,kammeraat2025correlated}.
		Reported correlation lengths span from cell-scale to tissue-scale, with correlation profiles showing either monotonic decay or negative correlation regimes depending on the underlying activity state. 
		WT (wild type) and BW (blebbistatin washout) conditions highlight how modulation of actomyosin contractility alters the spatial organization of velocity correlation.}
	\label{fig:velocity_corr_together}
\end{figure}

\subsubsection*{Each activity class leaves a distinct imprint on velocity correlation}

We compute the spatial velocity correlation,
\begin{equation}
	C(r)=\frac{\langle \mathbf{v}_i\cdot\mathbf{v}_j\rangle_{r}}{\langle |\mathbf{v}|^2\rangle},
\end{equation}
where $\langle \mathbf{v}_i\cdot\mathbf{v}_j\rangle_{r}$ denotes the average alignment between the velocities of cell pairs separated by a distance $r$, normalized by the mean squared cell velocity (see Methods).
Thus, $C(r)=1$ corresponds to perfectly aligned motion, $C(r)=0$ to uncorrelated motion, and $C(r)<0$ to oppositely directed motion.
We evaluate $C(r)$ for all four activity classes after the tissue reaches its circular steady state (Fig.~\ref{fig:velocity_corr_together}).

The correlation length, $\xi$, is obtained by exponentially fitting the positive branch of $C(r)$ prior to its first zero crossing (see Methods section for details).
The four activity classes produce qualitatively distinct correlation profiles, differing both in their correlation length and in the presence or absence of velocity anticorrelations (Fig.~\ref{fig:velocity_corr_together}).

AM exhibits the shortest-ranged correlations, with the velocity correlation function decaying rapidly to zero (Fig.~\ref{fig:velocity_corr_together} (a)) over a characteristic length scale of $\xi\sim0.5\sqrt{A_0}$ (Fig.~\ref{fig:velocity_corr_together} (b)).
This short correlation length reflects the spatially uncorrelated nature of vertex-level forcing, where active forces do not propagate beyond the local forcing scale.
As statistically independent forces cannot generate systematic velocity anti-alignment between spatially separated regions, $C(r)$ decays monotonically without developing a negative regime (Fig.~\ref{fig:velocity_corr_together} (a)).
PM likewise displays a monotonic exponential decay (Fig.~\ref{fig:velocity_corr_together} (a)), but with a noticeably larger correlation length ($\xi > \sqrt{A_0}$) (Fig.~\ref{fig:velocity_corr_together} (b)).
The polarized coherent forcing acting on each cell generates correlated motion over the cell scale before mechanical interactions progressively decorrelate the velocities.
This increased correlation length cleanly distinguishes PM from AM, revealing a difference that remained hidden in the structural and rheological observables.
As in AM, the decorrelation does not generate systematic anti-alignment: once polarity-induced coherence is lost, cell velocities become effectively independent, causing $C(r)$ to approach zero rather than develop a negative lobe (Fig.~\ref{fig:velocity_corr_together} (a)).\\
FC exhibits a rapid initial decay (Fig.~\ref{fig:velocity_corr_together} (a)) with $\xi<\sqrt{A_0}$ (Fig.~\ref{fig:velocity_corr_together} (b)), caused by the spatially unstructured contractility fluctuations that cannot synchronize motion beyond neighboring cells.
Unlike AM, however, $C(r)$ crosses zero and develops a pronounced negative lobe (Fig.~\ref{fig:velocity_corr_together} (a)). 
This anticorrelation arises from the kinematics of local contraction and extension: 
Cells on opposite sides of a locally contracting or expanding region move toward or away from one another, causing their center-of-mass velocities to become anti-aligned.
At larger separations, however, these local contractile domains no longer overlap, and distant cells experience statistically independent cellular movements, causing the anticorrelation to decay back toward zero.
Thus, although both AM and FC possess similarly short correlation lengths, the sign change in $C(r)$ provides a qualitative fingerprint unique to contractility-driven activity.\\
MR exhibits the most distinctive correlation profile.
The correlation decays much more gradually than in the other activity classes (Fig.~\ref{fig:velocity_corr_together} (a)), remaining positive over more than five cell diameters before developing a pronounced negative lobe, with the largest correlation length ($\xi\sim3\sqrt{A_0}$) of all four classes (Fig.~\ref{fig:velocity_corr_together} (a)).
This long-ranged structure is a direct consequence of the traveling Rho--ROCK--myosin waves (Fig.~\ref{fig:rho-rock-myosin}).
Cells within the same wavefront undergo contraction or expansion in phase and therefore move coherently, giving rise to extended positive correlations.
Cells separated by more than one wavelength occupy opposite phases of the wave and consequently move in opposing directions, producing the negative lobe.
The position of this lobe, therefore, provides a direct measure of the emergent wavelength of the underlying mechanochemical waves.

Experimental measurements of epithelial velocity correlations reveal a broad range of spatial organization in cell motion (Fig.~\ref{fig:velocity_corr_together}(c)).
NRK and MDCK monolayers show a monotonic decay of $C(r)$, with correlations extending over $\sim2$ and $\sim10$ cell sizes, respectively \cite{petitjean2010velocity}.
HBEC monolayers show correlations extending over $\sim7$--$10$ cell sizes \cite{garcia2015physics}.
In MDCK monolayers, the measurements of Thiagarajan \textit{et al.} show correlations extending to $\sim10$--$15$ cell sizes in the wild-type condition and to $\sim20$ cell sizes following blebbistatin washout, with the latter also exhibiting a pronounced negative correlation at intermediate distances \cite{thiagarajan2022pulsations}.
MDCK measurements by Kammeraat \textit{et al.} show correlations extending over $\sim5$--$7$ cell sizes \cite{kammeraat2025correlated}.
Thus, experimental epithelia exhibit a range of correlation lengths and both monotonic and non-monotonic correlation profiles.
Our results provide a new lens for interpreting this diversity, linking the spatial structure of velocity correlations to the underlying organization of cellular activity and suggesting a way to identify the mode of active force generation from cell motion.

\begin{figure}
	\centering
	\includegraphics[width=\linewidth]{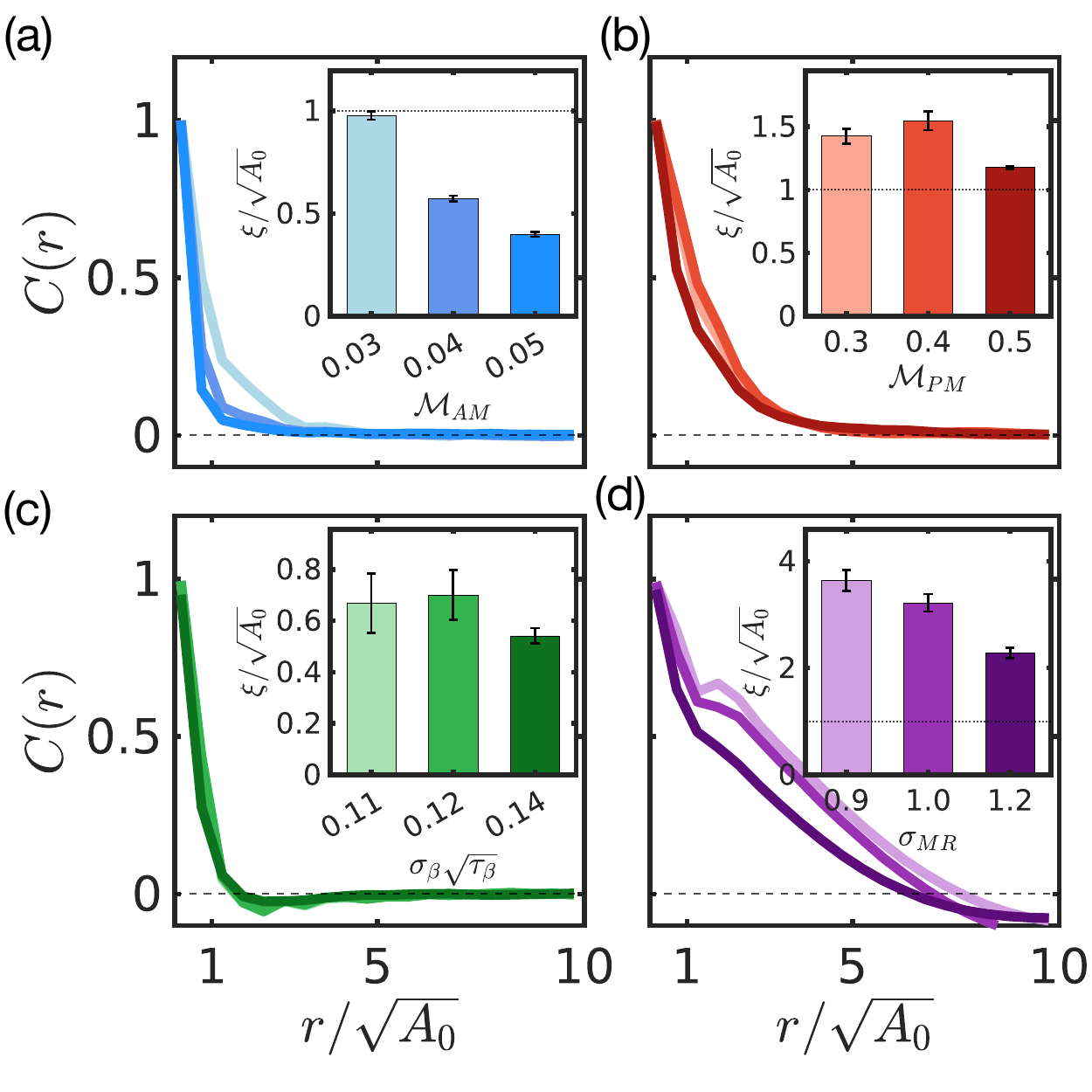}
	\caption{\textbf{Velocity correlation signatures across activity strengths.}
		(a) AM: Increasing the motility strength $\mathcal{M}_{\mathrm{AM}}$ accelerates the decay of $C(r)$, but correlations remain strictly positive at all activity levels, with $\xi$ (inset) consistently below one cell diameter.
		(b) PM: Across polar forcing strengths $\mathcal{M}_{\mathrm{PM}}$, $C(r)$ retains a monotonic exponential decay without a negative lobe, while $\xi$ (inset) remains above one cell diameter, distinguishing PM from AM.
		(c) FC: Varying the contractility fluctuation strength $\sigma_\beta\sqrt{\tau_\beta}$ preserves the short-ranged decay and shallow anticorrelation of $C(r)$, with $\xi$ (inset) remaining below one cell diameter. 
		The negative lobe reflects the spatial organization of contractile fluctuations rather than their magnitude.
		(d) MR: Changes in the coupling noise strength $\sigma_{\mathrm{MR}}$ preserve the long-ranged positive correlations and pronounced negative lobe associated with mechanochemical waves, with $\xi$ (inset) remaining the largest among all activity classes.
	}
	\label{fig:velocity_corr_seperate}
\end{figure}

\subsubsection*{Velocity correlation signatures are robust to activity strength}
In experiments, cellular activity can be tuned over a range of strengths by modifying motility, contractility, or signaling pathways, for example, through changes in substrate conditions, myosin inhibition or washout, or perturbations of signaling pathways that regulate force generation.
We therefore ask whether the velocity correlation signatures identified above remain robust when the activity strength is varied within each class (Fig.~\ref{fig:velocity_corr_seperate}).

Increasing AM strength enhances local velocity fluctuations, leading to faster decay of $C(r)$, correlations remain short-ranged ($\xi<\sqrt{A_0}$) and strictly positive across all conditions (Fig.~\ref{fig:velocity_corr_seperate}(a)).
PM, in contrast, preserves long-ranged positive correlations ($\xi>\sqrt{A_0}$),  reflecting persistent cell-scale coordination over a broad range of polar activity strengths (Fig.~\ref{fig:velocity_corr_seperate}(b)).
FC consistently exhibits short-ranged correlations with a negative lobe, indicating that local contractile fluctuations generate anti-correlated flows independent of fluctuation amplitude (Fig.~\ref{fig:velocity_corr_seperate}(c)).
MR remains uniquely characterized by long-ranged correlations followed by a pronounced negative lobe, reflecting the spatial organization imposed by mechanochemical waves (Fig.~\ref{fig:velocity_corr_seperate}(d)).

Thus, varying the magnitude of activity primarily alters the strength and decay scale of velocity correlations while preserving the qualitative structure of $C(r)$ associated with the underlying mechanism. 
This suggests that velocity correlations can serve as an experimentally accessible discriminator of activity modes, allowing the physical origin of cellular activity to be inferred from live-cell imaging without direct measurement of molecular activity.

\section*{Discussion}

This work presents a framework to infer the underlying mode of active force generation from macroscopic measurements by unifying diverse forms of epithelial activity into a common physical framework grounded in the origin and spatial organization of active force generation. By comparing the cellular dynamics across different activity generation mechanisms---localized filopodial force generation by stochastic membrane fluctuations (AM), coherent cell polarity (PM), fluctuating cortical contractility (FC), and mechanochemical regulation (MR)---mechanisms previously studied separately \cite{islam2025motility, drubin1996origins, curran2017myosin, koride2014mechanochemical}, we show that each occupies a distinct region of a broader biophysical landscape. 
Although all four activity classes drive tissue fluidization and converge to similar macroscopic morphologies, they do so through distinct microscopic pathways. 
These differences propagate across scales, from junctional rearrangements and cell dynamics to tissue structure, rheology, and collective motion, demonstrating that fluidization is not a unique mechanical state but a common outcome reached through different modes of force generation.
This perspective shifts the focus from individual model implementations toward understanding how the physical origin and organization of active forces shape tissue remodeling across scales.

Importantly, we also identify experimentally accessible physical signatures that 
distinguish the four activity classes within our model framework.
While structural descriptors distinguish motility-driven from contractility-driven tissues, and rheological measurements reveal characteristic signatures of mechanochemical feedback, neither alone completely resolves the underlying mode of activity.
Spatial velocity correlations, by contrast, simultaneously encode both the range and spatial organization of collective motion, providing a sensitive observable that can distinguish activity classes that otherwise produce similar structural and mechanical outcomes within the model framework.
Since these correlations can be extracted directly from standard live-cell imaging through particle image velocimetry \cite{petitjean2010velocity, garcia2015physics, thiagarajan2022pulsations, angelini2011glass} or cell-tracking approaches \cite{mehes2012collective,mohammed2019substrate}, they offer a practical route for constraining the dominant mode of activity without requiring mechanical perturbations or molecular reporters. 

The framework also provides a lens for interpreting experimentally observed velocity correlations across diverse epithelial systems. 
Long-ranged positive correlations without a negative regime, as observed during collective migration and tissue expansion, are consistent with coordinated motion dominated by coherent polarity-driven motion or large-scale flows, resembling the PM class \cite{petitjean2010velocity, kammeraat2025correlated}. 
However, long correlation lengths alone do not uniquely identify the underlying mechanism, as coordinated contractile dynamics can generate similarly extended coherence. 
For example, pulsatile actomyosin-driven dynamics exhibit large-scale correlated motion consistent with the MR class, while blebbistatin washout further demonstrates that increased coherence can be accompanied by anti-correlated flows, highlighting the mechanistic information contained in the full shape of $C(r)$ \cite{thiagarajan2022pulsations}. 
Short-ranged correlations with a negative lobe are consistent with locally organized contractile fluctuations, as captured by FC, whereas the progressive loss of collective motion in aging HBEC monolayers leads to cell-scale, monotonic correlations without anticorrelations, resembling the AM regime \cite{garcia2015physics}. 
Together, these observations suggest that correlation length and the presence or absence of velocity anticorrelations provide complementary constraints on the physical origin of epithelial activity.

The activity classes studied here are distinguished primarily by the spatial scale at which forces are generated and coordinated, but the temporal organization of these forces --- their persistence and temporal correlation --- remains an important dimension that has yet to be explored systematically.
A natural extension is to introduce finite persistence times into polar motility, thereby connecting this framework to self-propelled particle models of epithelial fluidity \cite{bi2016motility}.
Also, living epithelia rarely operate through a single mechanism of activity in isolation, and extending this framework to mixed activity states will provide insight into how different modes of force generation interact during tissue remodeling. 
Further incorporating extracellular matrix interactions and three-dimensional architectures will strengthen this framework's applicability to developmental and pathological systems.

Taken together, our results establish a connection between microscopic modes of force generation and macroscopic tissue behavior.
By showing that distinct modes of activity leave characteristic signatures across cellular rearrangements, rheology, and collective motion, we provide an experimentally accessible framework for linking measurable tissue dynamics to the physical organization of active force generation in living epithelia.

\section*{Methods}
\subsection*{Parameters} {\label{SI_sec:parameter}}
Fixed simulation parameters used are $\eta=1$, $\lambda=1$, $A_0=1$, $\gamma=0.06$, and $\beta=0.05$ (unless coupled to Myosin dynamics). 
All quantities are nondimensionalized using $\sqrt{A_0}$ as the length scale and $\eta/(\lambda A_0)$ as the time scale; although no separate notation is introduced for nondimensional variables, all variables below should be understood in these units unless stated otherwise.
The deterministic Euler timestep was $dt=10^{-4}$, with T1 and T2 thresholds set to $0.1\sqrt{A_0}$ and $0.25A_0$, respectively.\\
For AM, $\mathcal{M}_{\mathrm{AM}}=0.04$, corresponding to vertex displacements that decorrelate on timescales comparable to mechanical relaxation and to effective speeds of $0.1-1~\mu\mathrm{m/min}$.
PM was set to $\mathcal{M}_{\mathrm{polar}}\sim0.5$, corresponding to persistent motion over a cell size within one relaxation time, causing speeds of $\sim1-10~\mu\mathrm{m/min}$.\\
For FC, activity was introduced through Ornstein--Uhlenbeck fluctuations with $\tau_\beta=1$ and $\sigma_\beta=0.12$ around $\beta_0=0.05$.
Here, $\tau_\beta$ sets myosin correlation time relative to mechanical relaxation, while $\sigma_\beta$ controls the magnitude of transient tension fluctuations.
Together, $\sigma_\beta\sqrt{\tau_\beta}$ determines dimensionless activity parameter for FC.\\
For MR, the Rho--ROCK--Myosin ODE parameters are listed in Table~\ref{tab:mr_parameters}.
The coupling between myosin activity and cellular contractility is assumed linear with strength $\alpha=0.1$, with additive coupling fluctuations of strength $\sigma_{\mathrm{MR}} = 1.0$ capturing variability in the transmission of biochemical activity into mechanical contractility.
\begin{table}[h!]
	\centering
	\caption{Parameters for the Rho--ROCK--Myosin signaling dynamics in the mechanochemical regulation (MR) model.}
	\begin{tabular}{lll}
		\hline
		\textbf{Parameter} & \textbf{Symbol} & \textbf{Value} \\
		\hline
		Hill coefficient           & $n$            & $2.0$ \\
		Activation threshold       & $K$            & $0.01$ \\
		Rho activation rate        & $A_{\rho}$     & $10.0$ \\
		ROCK activation rate       & $A_{R}$        & $7.0$ \\
		Myosin activation rate     & $A_{M}$        & $5.0$ \\
		Rho decay rate             & $D_{\rho}$     & $0.1$ \\
		ROCK decay rate            & $D_{R}$        & $0.1$ \\
		Myosin decay rate          & $D_{M}$        & $0.1$ \\
		\hline
	\end{tabular}
	\label{tab:mr_parameters}
\end{table}

\subsection*{Analysis}
% \subsubsection*{Mean Squared displacement}
% The mean square displacement for the tissue is defined as, 

% \begin{equation}
	%     \mathrm{MSD}(t) =  \langle |\mathbf{r}(t) - \mathbf{r}(0)|^2 \rangle  
	% \end{equation}
% where $\mathbf{r}(t)$ is the center of mass of a cell (at time $t$) calculated from the mean of the vertex positions of the cell.

\subsubsection*{Stress Tensor}
The stress tensor for an individual cell is calculated by, 
\begin{equation}
	\boldsymbol{\hat{\sigma}_c} = - \Pi_c \hat{\mathbf I} + \frac{1}{2 A_c} \sum_{e \in c} \boldsymbol{T}_e \otimes \boldsymbol{l}_e
\end{equation}
Where $ \Pi_c = - \frac{\partial U_c}{\partial A_c}$ is the hydrostatic pressure and $\boldsymbol{T}_e = \frac{\partial U_c }{\partial \boldsymbol{l}_e}$ is the line tension (or shear stress ) term.

\subsubsection*{Stress Relaxation Experiment}
A stress relaxation experiment is performed by applying an affine shear deformation,
${\bf \gamma}(t)=\left(\begin{array}{cc}1 & \epsilon_{xy} \\ 0 & 1\end{array}\right)$,
and holding the tissue boundaries at fixed strain. 
The resulting bulk shear stress relaxation is measured away from the boundaries.

\subsubsection*{Spatial Velocity Correlation Function}

At each sampled time $t$, the instantaneous velocity of cell $i$ is estimated as

\begin{equation}
	\mathbf{v}_i(t)=\frac{\mathbf{r}_i(t+\delta t)-\mathbf{r}_i(t)}{\delta t},
\end{equation}

where $\mathbf{r}_i(t)$ is the centroid of cell $i$, obtained by averaging its vertex positions, and $\delta t$ is chosen small enough to resolve cell dynamics while suppressing numerical noise.
Only cells present in both ($t$ and $t + \delta t$) snapshots contribute to the velocity estimate.\\
The spatial velocity correlation function is implemented as,

\begin{equation}
	C(r)=\frac{\displaystyle \overline{\left\langle \mathbf{v}_i\cdot\mathbf{v}_j\right\rangle_{|r_{ij}-r|<\Delta r/2}}}
	{\overline{|\mathbf{v}|^2}},
\end{equation}

where $r_{ij}=|\mathbf{r}_i-\mathbf{r}_j|$, the angular brackets denote averaging over all cell pairs within the distance bin centered at $r$, and the overbar denotes time averaging over steady-state snapshots.
The denominator is the mean squared speed over all cells and time windows, providing a fixed normalization that makes $C(r)$ dimensionless and comparable across activity classes.
The average includes all cells simultaneously present between consecutive snapshots.\\
% The $r=0$ value is obtained by including self-pairs ($i=j$), giving $C(0) \to 1$ by construction and providing the anchor for correlation-length extraction.
% Since the tissue evolves under open boundary conditions, pair distances are computed using raw Euclidean distances without periodic correction.
The correlation length $\xi$ is obtained by fitting the positive region of $C(r)$ to, 

\begin{equation}
	C(r)\approx C_0\exp(-r/\xi),
\end{equation}
using nonlinear least squares.
% Fits are restricted to $r>0$ and $C(r)>0$ to exclude the negative lobe observed for some activity classes.
The uncertainty in $\xi$ is reported as the standard error from the nonlinear fit.

\subsection*{Author Contribution}
M.S.R. and A.G. conceived the project. M.S.R. and A.G. supervised the project.
S.I. performed research and simulations.
S.I., M.S.R., and A.G. designed analysis methods. 
S.I. performed all the analyses.
S.I., M.S.R., and A.G. wrote the paper jointly.

\section*{Acknowledgments}

All simulations were performed on the \textit{Paramseva} supercomputers under the \textit{National Supercomputing Mission, India}, the \textit{IITH Kanad cluster}, and a local high-performance workstation.\\
S.I. acknowledges the Prime Minister Research Fellowship (PMRF, ID-2002732) for financial support through a research fellowship. A. G. acknowledges SERB-DST (India) Projects MTR/2022/000232, CRG/2023/007056-G, DST (India) grant no. \\
DST/NSM/R\&D\_HPC\_Applications/2021/05 and grant no.
SR/FST/PSI-215/2016, and 
IITH for Seed Grant No. \\IITH/2020/09.
M.S.R. acknowledges the SERB (India) project SRG/2021/001020 and 
IIT Hyderabad for financial support.

% Uncomment if using bibtex (default)
\bibliography{Citations}

@article{islam2025motility,
  title={Motility-Driven Viscoelastic Control of Tissue Morphology in Presomitic Mesoderm},
  author={Islam, Sahil and Rizvi, Mohd and Gupta, Anupam and others},
  journal={arXiv preprint arXiv:2510.24314},
  year={2025}
}

@article{heisenberg2013forces,
  title={Forces in tissue morphogenesis and patterning},
  author={Heisenberg, Carl-Philipp and Bella{\"\i}che, Yohanns},
  journal={Cell},
  volume={153},
  number={5},
  pages={948--962},
  year={2013},
  publisher={Elsevier}
}

@article{kim2021embryonic,
  title={Embryonic tissues as active foams},
  author={Kim, Sangwoo and Pochitaloff, Marie and Stooke-Vaughan, Georgina A and Camp{\`a}s, Otger},
  journal={Nature physics},
  volume={17},
  number={7},
  pages={859--866},
  year={2021},
  publisher={Nature Publishing Group UK London}
}

@article{bi2015density,
  title={A density-independent rigidity transition in biological tissues},
  author={Bi, Dapeng and Lopez, JH and Schwarz, Jennifer M and Manning, M Lisa},
  journal={Nature Physics},
  volume={11},
  number={12},
  pages={1074--1079},
  year={2015},
  publisher={Nature Publishing Group UK London}
}

@article{bi2016motility,
  title={Motility-driven glass and jamming transitions in biological tissues},
  author={Bi, Dapeng and Yang, Xingbo and Marchetti, M Cristina and Manning, M Lisa},
  journal={Physical Review X},
  volume={6},
  number={2},
  pages={021011},
  year={2016},
  publisher={APS}
}

@incollection{honda2022vertex,
  title={Vertex model},
  author={Honda, Hisao and Nagai, Tatsuzo},
  booktitle={Mathematical models of cell-based morphogenesis: passive and active remodeling},
  pages={39--57},
  year={2022},
  publisher={Springer}
}

@article{farhadifar2007influence,
  title={The influence of cell mechanics, cell-cell interactions, and proliferation on epithelial packing},
  author={Farhadifar, Reza and R{\"o}per, Jens-Christian and Aigouy, Benoit and Eaton, Suzanne and J{\"u}licher, Frank},
  journal={Current biology},
  volume={17},
  number={24},
  pages={2095--2104},
  year={2007},
  publisher={Elsevier}
}

@article{lin2018dynamic,
  title={Dynamic migration modes of collective cells},
  author={Lin, Shao-Zhen and Ye, Sang and Xu, Guang-Kui and Li, Bo and Feng, Xi-Qiao},
  journal={Biophysical journal},
  volume={115},
  number={9},
  pages={1826--1835},
  year={2018},
  publisher={Elsevier}
}

@article{zimmermann2014formation,
  title={Formation of transient lamellipodia},
  author={Zimmermann, Juliane and Falcke, Martin},
  journal={PLoS one},
  volume={9},
  number={2},
  pages={e87638},
  year={2014},
  publisher={Public Library of Science San Francisco, USA}
}

@article{zimmermann2013existence,
  title={On the existence and strength of stable membrane protrusions},
  author={Zimmermann, Juliane and Falcke, Martin},
  journal={New Journal of Physics},
  volume={15},
  number={1},
  pages={015021},
  year={2013},
  publisher={IOP Publishing}
}

@article{ji2008fluctuations,
  title={Fluctuations of intracellular forces during cell protrusion},
  author={Ji, Lin and Lim, James and Danuser, Gaudenz},
  journal={Nature cell biology},
  volume={10},
  number={12},
  pages={1393--1400},
  year={2008},
  publisher={Nature Publishing Group UK London}
}

@article{machacek2006morphodynamic,
  title={Morphodynamic profiling of protrusion phenotypes},
  author={Machacek, Matthias and Danuser, Gaudenz},
  journal={Biophysical journal},
  volume={90},
  number={4},
  pages={1439--1452},
  year={2006},
  publisher={Elsevier}
}

@article{ridley2003cell,
  title={Cell migration: integrating signals from front to back},
  author={Ridley, Anne J and Schwartz, Martin A and Burridge, Keith and Firtel, Richard A and Ginsberg, Mark H and Borisy, Gary and Parsons, J Thomas and Horwitz, Alan Rick},
  journal={Science},
  volume={302},
  number={5651},
  pages={1704--1709},
  year={2003},
  publisher={American Association for the Advancement of Science}
}

@article{petrie2009random,
  title={Random versus directionally persistent cell migration},
  author={Petrie, Ryan J and Doyle, Andrew D and Yamada, Kenneth M},
  journal={Nature reviews Molecular cell biology},
  volume={10},
  number={8},
  pages={538--549},
  year={2009},
  publisher={Nature Publishing Group UK London}
}

@article{drubin1996origins,
  title={Origins of cell polarity},
  author={Drubin, David G and Nelson, W James},
  journal={Cell},
  volume={84},
  number={3},
  pages={335--344},
  year={1996},
  publisher={Elsevier}
}

@article{murrell2015forcing,
  title={Forcing cells into shape: the mechanics of actomyosin contractility},
  author={Murrell, Michael and Oakes, Patrick W and Lenz, Martin and Gardel, Margaret L},
  journal={Nature reviews Molecular cell biology},
  volume={16},
  number={8},
  pages={486--498},
  year={2015},
  publisher={Nature Publishing Group UK London}
}

@article{curran2017myosin,
  title={Myosin II controls junction fluctuations to guide epithelial tissue ordering},
  author={Curran, Scott and Strandkvist, Charlotte and Bathmann, Jasper and De Gennes, Marc and Kabla, Alexandre and Salbreux, Guillaume and Baum, Buzz},
  journal={Developmental cell},
  volume={43},
  number={4},
  pages={480--492},
  year={2017},
}

@article{koride2014mechanochemical,
  title={Mechanochemical regulation of oscillatory follicle cell dynamics in the developing Drosophila egg chamber},
  author={Koride, Sarita and He, Li and Xiong, Li-Ping and Lan, Ganhui and Montell, Denise J and Sun, Sean X},
  journal={Molecular biology of the cell},
  volume={25},
  number={22},
  pages={3709--3716},
  year={2014},
  publisher={The American Society for Cell Biology}
}

@article{koride2018epithelial,
  title={Epithelial vertex models with active biochemical regulation of contractility can explain organized collective cell motility},
  author={Koride, Sarita and Loza, Andrew J and Sun, Sean X},
  journal={APL bioengineering},
  volume={2},
  number={3},
  year={2018},
  publisher={AIP Publishing}
}

@article{michaut2025extracellular,
  title={Extracellular volume expansion drives vertebrate axis elongation},
  author={Michaut, Arthur and Mongera, Alessandro and Gupta, Anupam and Tarazona, Oscar A and Serra, Mattia and Kefala, Georgia-Maria and Rigoni, Pietro and Lee, Jong Gwan and Rivas, Felipe and Hall, Adam R and others},
  journal={Current Biology},
  volume={35},
  number={4},
  pages={843--853},
  year={2025},
  publisher={Elsevier}
}

@article{staple2010mechanics,
  title={Mechanics and remodelling of cell packings in epithelia},
  author={Staple, Douglas B and Farhadifar, Reza and R{\"o}per, J-C and Aigouy, Benoit and Eaton, Suzanne and J{\"u}licher, Frank},
  journal={The European Physical Journal E},
  volume={33},
  number={2},
  pages={117--127},
  year={2010},
  publisher={Springer}
}

@article{papafilippou2025interplay,
  title={Interplay of damage and repair in the control of epithelial tissue integrity in response to cyclic loading},
  author={Papafilippou, Eleni and Baldauf, Lucia and Charras, Guillaume and Kabla, Alexandre J and Bonfanti, Alessandra},
  journal={Current Opinion in Cell Biology},
  volume={94},
  pages={102511},
  year={2025},
  publisher={Elsevier}
}

@article{petitjean2010velocity,
  title={Velocity fields in a collectively migrating epithelium},
  author={Petitjean, Laurence and Reffay, Myriam and Grasland-Mongrain, Erwan and Poujade, Mathieu and Ladoux, Beno{\i}t and Buguin, Axel and Silberzan, Pascal},
  journal={Biophysical journal},
  volume={98},
  number={9},
  pages={1790--1800},
  year={2010},
  publisher={Elsevier}
}

@article{garcia2015physics,
  title={Physics of active jamming during collective cellular motion in a monolayer},
  author={Garcia, Simon and Hannezo, Edouard and Elgeti, Jens and Joanny, Jean-Fran{\c{c}}ois and Silberzan, Pascal and Gov, Nir S},
  journal={Proceedings of the National Academy of Sciences},
  volume={112},
  number={50},
  pages={15314--15319},
  year={2015},
  publisher={National Academy of Sciences}
}

@article{thiagarajan2022pulsations,
  title={Pulsations and flows in tissues as two collective dynamics with simple cellular rules},
  author={Thiagarajan, Raghavan and Bhat, Alka and Salbreux, Guillaume and Inamdar, Mandar M and Riveline, Daniel},
  journal={Iscience},
  volume={25},
  number={10},
  year={2022},
  publisher={Elsevier}
}

@article{kammeraat2025correlated,
  title={Correlated cell movements drive epithelial finger formation},
  author={Kammeraat, Sander C and Keta, Yann-Edwin and Appleton, Paul and Newton, Ian P and Liverpool, Tanniemola B and Sknepnek, Rastko and N{\"a}thke, Inke and Henkes, Silke},
  journal={arXiv preprint arXiv:2508.01046},
  year={2025}
}

@article{guillot2013mechanics,
  title={Mechanics of epithelial tissue homeostasis and morphogenesis},
  author={Guillot, Charl{\`e}ne and Lecuit, Thomas},
  journal={Science},
  volume={340},
  number={6137},
  pages={1185--1189},
  year={2013},
  publisher={American Association for the Advancement of Science}
}

@article{davidson2012epithelial,
  title={Epithelial machines that shape the embryo},
  author={Davidson, Lance A},
  journal={Trends in Cell Biology},
  volume={22},
  number={2},
  pages={82--87},
  year={2012},
  publisher={Elsevier}
}

@article{pena2024cellular,
  title={Cellular and molecular mechanisms of skin wound healing},
  author={Pe{\~n}a, Oscar A and Martin, Paul},
  journal={Nature Reviews Molecular Cell Biology},
  volume={25},
  number={8},
  pages={599--616},
  year={2024},
  publisher={Nature Publishing Group UK London}
}

@article{macara2014epithelial,
  title={Epithelial homeostasis},
  author={Macara, Ian G and Guyer, Richard and Richardson, Graham and Huo, Yongliang and Ahmed, Syed M},
  journal={Current biology},
  volume={24},
  number={17},
  pages={R815--R825},
  year={2014},
  publisher={Elsevier}
}

@article{thiery2009epithelial,
  title={Epithelial-mesenchymal transitions in development and disease},
  author={Thiery, Jean Paul and Acloque, Herv{\'e} and Huang, Ruby YJ and Nieto, M Angela},
  journal={cell},
  volume={139},
  number={5},
  pages={871--890},
  year={2009},
  publisher={Elsevier}
}

@article{marchiando2010epithelial,
  title={Epithelial barriers in homeostasis and disease},
  author={Marchiando, Amanda M and Graham, W Vallen and Turner, Jerrold R},
  journal={Annual Review of Pathology: Mechanisms of Disease},
  volume={5},
  number={1},
  pages={119--144},
  year={2010},
  publisher={Annual Reviews}
}

@article{vasquez2016force,
  title={Force transmission in epithelial tissues},
  author={Vasquez, Claudia G and Martin, Adam C},
  journal={Developmental Dynamics},
  volume={245},
  number={3},
  pages={361--371},
  year={2016},
  publisher={Wiley Online Library}
}

@article{kumar2025forces,
  title={Forces at the scale of the cell},
  author={Kumar, K Vijay and Inamdar, Mandar M and Pullarkat, Pramod A and Menon, Gautam I},
  journal={arXiv preprint arXiv:2512.08311},
  year={2025}
}

@article{birchmeier1996epithelial,
  title={Epithelial-mesenchymal transitions in cancer progression},
  author={Birchmeier, Carmen and Birchmeier, Walter and Brand-Saberi, Beate},
  journal={Cells Tissues Organs},
  volume={156},
  number={3},
  pages={217--226},
  year={1996},
  publisher={S. Karger AG Basel, Switzerland}
}

@article{amano2010rho,
  title={Rho-kinase/ROCK: a key regulator of the cytoskeleton and cell polarity},
  author={Amano, Mutsuki and Nakayama, Masanori and Kaibuchi, Kozo},
  journal={Cytoskeleton},
  volume={67},
  number={9},
  pages={545--554},
  year={2010},
  publisher={Wiley Online Library}
}

@article{janshoff2021viscoelastic,
  title={Viscoelastic properties of epithelial cells},
  author={Janshoff, Andreas},
  journal={Biochemical society transactions},
  volume={49},
  number={6},
  pages={2687--2695},
  year={2021},
  publisher={Portland Press Ltd.}
}

@article{style2014traction,
  title={Traction force microscopy in physics and biology},
  author={Style, Robert W and Boltyanskiy, Rostislav and German, Guy K and Hyland, Callen and MacMinn, Christopher W and Mertz, Aaron F and Wilen, Larry A and Xu, Ye and Dufresne, Eric R},
  journal={Soft matter},
  volume={10},
  number={23},
  pages={4047--4055},
  year={2014},
  publisher={Royal Society of Chemistry}
}

@article{safa2024active,
  title={Active viscoelastic models for cell and tissue mechanics},
  author={Safa, Bahareh Tajvidi and Huang, Changjin and Kabla, Alexandre and Yang, Ruiguo},
  journal={Royal Society open science},
  volume={11},
  number={4},
  pages={231074},
  year={2024}
}

@article{papafilippou2026emergent,
  title={Emergent Intercellular Junction Stability during Cyclic Tissue Loading},
  author={Papafilippou, Eleni and Bonfanti, Alessandra and Charras, Guillaume and Kabla, Alexandre},
  journal={Biophysical Journal},
  year={2026},
  publisher={Elsevier}
}

@article{angelini2011glass,
  title={Glass-like dynamics of collective cell migration},
  author={Angelini, Thomas E and Hannezo, Edouard and Trepat, Xavier and Marquez, Manuel and Fredberg, Jeffrey J and Weitz, David A},
  journal={Proceedings of the National Academy of Sciences},
  volume={108},
  number={12},
  pages={4714--4719},
  year={2011},
  publisher={National Academy of Sciences}
}

@article{mehes2012collective,
  title={Collective motion of cells mediates segregation and pattern formation in co-cultures},
  author={M{\'e}hes, El{\H{o}}d and Mones, Enys and Nemeth, Valeria and Vicsek, Tamas},
  journal={PloS one},
  volume={7},
  number={2},
  pages={e31711},
  year={2012},
  publisher={Public Library of Science San Francisco, USA}
}

@article{mohammed2019substrate,
  title={Substrate area confinement is a key determinant of cell velocity in collective migration},
  author={Mohammed, Danahe and Charras, Guillaume and Vercruysse, El{\'e}onore and Versaevel, Marie and Lantoine, Jos{\'e}phine and Alaimo, Laura and Bruy{\`e}re, C{\'e}line and Luciano, Marine and Glinel, Karine and Delhaye, Geoffrey and others},
  journal={Nature Physics},
  volume={15},
  number={8},
  pages={858--866},
  year={2019},
  publisher={Nature Publishing Group UK London}
}

% Uncomment if using biblatex
% \printbibliography

%\section*{Supplementary Material}
%
%An online supplement to this article can be found by visiting BJ Online at \url{http://www.biophysj.org}.

%%---------------------------------------------------------
%%---------------------------------------------------------
%%---------------------------------------------------------

\clearpage

% Figures, tables, equations and pages in the supplement are numbered S1, S2 etc.
\renewcommand{\thefigure}{S\arabic{figure}}
\renewcommand{\thetable}{S\arabic{table}}
\renewcommand{\theequation}{S\arabic{equation}}
\renewcommand{\thepage}{S\arabic{page}}

\setcounter{figure}{0}
\setcounter{table}{0}
\setcounter{equation}{0}
\setcounter{page}{1}

%\onecolumngrid

\onecolumngrid
\begin{center}
		
		{\LARGE \textbf{Supplementary Information \\ {\Large for}}}\\[1em]
		
		{\Large \textbf{{\mstitle}
		}}\\[1em]
		
		Sahil Islam$^{1}$,
		Anupam Gupta$^{1}$,
		Mohd. Suhail Rizvi$^{2}$,
		\\[1em]
		
		{\small
			$^{1}$ Department of Physics, Indian Institute of Technology Hyderabad, Telangana, India\\
			$^{2}$ Department of Biomedical Engineering, Indian Institute of Technology Hyderabad, Telangana, India\\
		}
		
		\vspace{1em}
		
		%{\small Corresponding author: agupta@phy.iith.ac.in}
		
\end{center}

\twocolumngrid

\begin{figure}
	\centering
	\includegraphics[width=\linewidth]{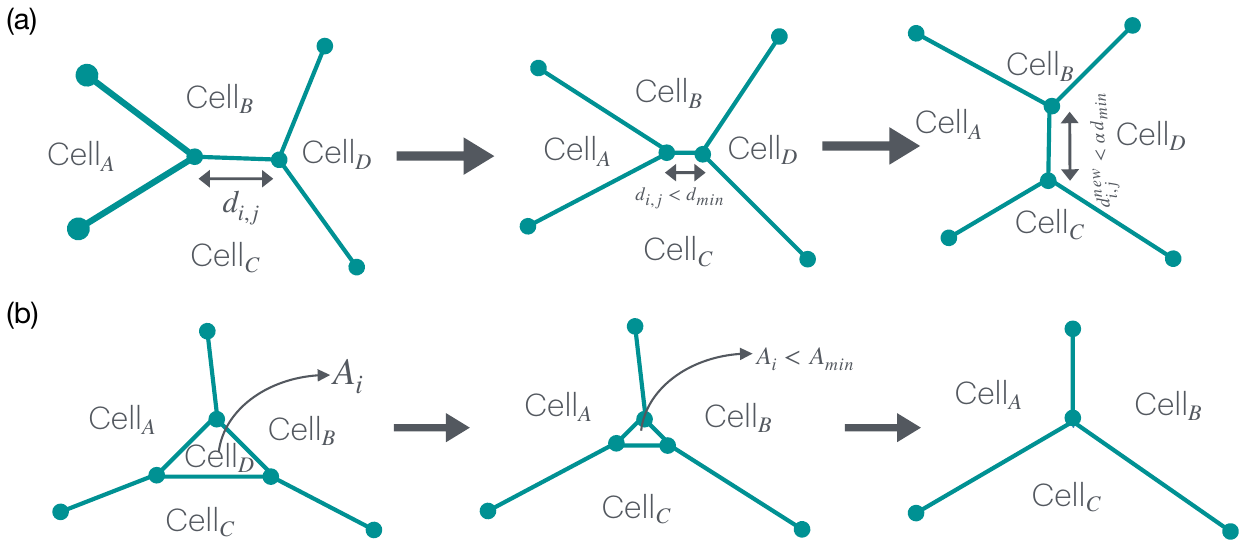}
	\caption{\textbf{Schematic illustration of topological rearrangements in the vertex model.}
		(a) Schematic representing T1 transition in which an existing cell--cell edge shrinks to zero length and is replaced by a new edge connecting the previously non-neighboring cells. 
		(b) Schematic showing T2 transition in which if a cell shrinks below a critical area it is removed from the tissue, followed by a topological rearrangement of the surrounding junctions.}
	\label{fig:t1t2}
\end{figure}

\begin{figure}
	\centering
	\includegraphics[width=\linewidth]{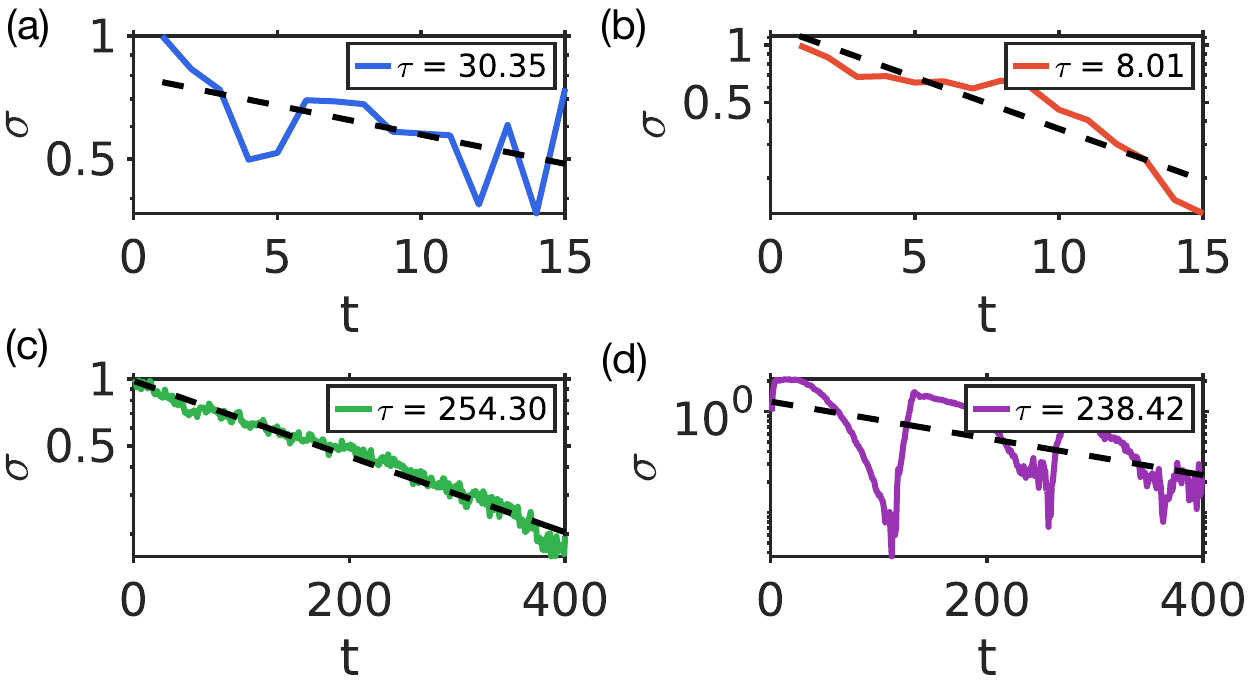}
	\caption{\textbf{Linear fitting of stress relaxation curves in a lin-log scale showing exponential decay behavior.
	}}
	\label{fig:stress_fitting}
\end{figure}

\begin{figure}
	\centering
	\includegraphics[width=\linewidth]{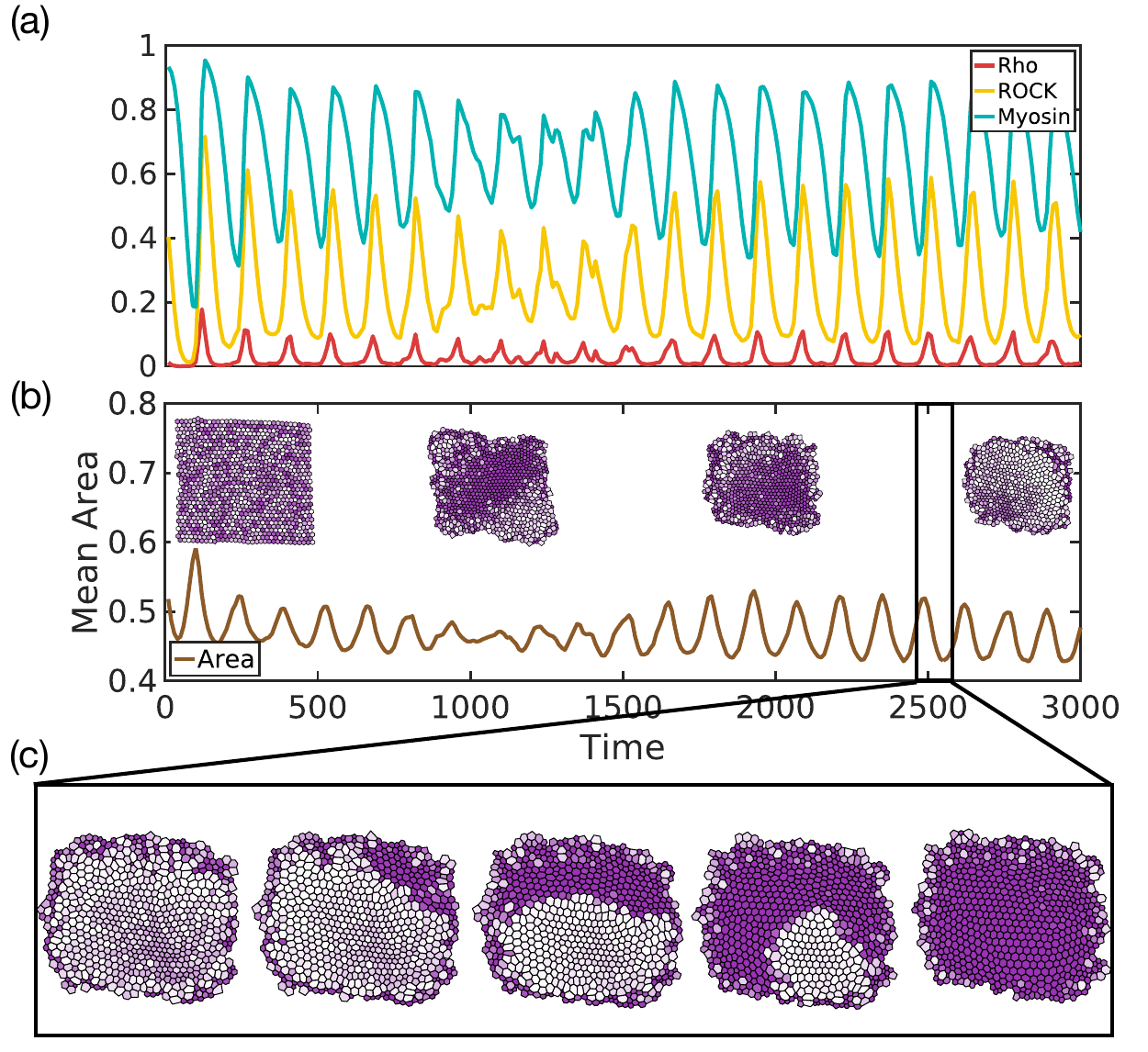}
	\caption{\textbf{Spatiotemporal dynamics of Rho--ROCK signaling and myosin activity.}\\
		(a) Time evolution of the spatially averaged Rho--ROCK and myosin concentrations, showing sustained oscillatory dynamics.
		(b) Oscillations in the mean tissue area during the evolution toward the circular steady state.
		The inset shows the corresponding tissue morphology.
		(c) Propagation of a traveling myosin wave across the tissue over a selected time interval.
		\label{fig:rho-rock-myosin}}
\end{figure}

\end{document}